\documentclass[runningheads]{llncs}

\usepackage[boxed]{algorithm}
\usepackage[noend]{algorithmic}
\usepackage{epsfig}
\usepackage{latexsym}
\usepackage{longtable}
\usepackage{psboxit}
\usepackage{boxedminipage}
\floatstyle{ruled}
\newfloat{ex05}{h}{lox}
\floatname{ex05}{Ex.}
\def\dfn#1{{\em #1}}

\def\sicstus{SICStus}
\def\tracesize{\footnotesize}
\newcommand{\caplab}[2]{\caption{\label{#1} #2}}

\newcommand{\tabbegin}[1]{\begin{table*}[#1]\centering}
\newcommand{\tabend}{\end{table*}}
\newcommand{\figbegin}[1]{\begin{figure*}[#1]}
\newcommand{\figend}[2]{\caplab{#1}{#2}\end{figure*}}

\newcommand{\hlineskip}{\noalign{\smallskip}\hline\noalign{\smallskip}}
\newcommand{\tabhead}[1]{\textbf{#1}}
\newcommand{\evtabbegin}[3]{\begin{minipage}{\linewidth}
                        \begin{tabular}{p{#1cm}|p{#2cm}|p{-#3cm + 
                        \linewidth}}
                        \hline\noalign{\smallskip} 
                        \multicolumn{1}{l}{\tabhead{Trace Event}} & 
                        \multicolumn{1}{l}{\tabhead{Attributes}} & 
                        \multicolumn{1}{l}{\tabhead{Meaning}} \\
                        \noalign{\smallskip} \hline \noalign{\smallskip}}
\newcommand{\evtabend}{\end{tabular}\end{minipage}}
\newcommand{\standout}[1]{\textbf{#1}}
\newcommand{\greybox}[1]{\psboxit{box .7 setgray fill}{#1}}

\newcommand{\ttitem}[1]{\item{\texttt{#1}}}

\newcommand{\mtt}[1]{$\mathtt{#1}$} 
\newcommand{\mexp}[1]{$#1$}

 \newcommand{\clpfd}{CLP(FD)}

\newenvironment{quotetab} {\begin{quote}\begin{tabbing}}
{\end{tabbing}\end{quote}}

\begin{document}
\setcounter{page}{1}
\title{Tracing and Explaining Execution of \clpfd\ Programs}
\titlerunning{Tracing and Explaining Execution of \clpfd\ Programs}
\author{Magnus {\AA}gren \and Tam{\'a}s Szeredi \and Nicolas Beldiceanu \and Mats Carlsson}
\authorrunning{M. {\AA}gren, T. Szeredi, N. Beldiceanu, M. Carlsson}
\institute{SICS, L{\"a}gerhyddsv. 18, SE-752~37~~UPPSALA, Sweden\\
        \email{\symbol{`\{}magren,tszeredi,nicolas,matsc\symbol{`\}}@sics.se}}

\maketitle

\addtocounter{footnote}{1}
\footnotetext{In Alexandre Tessier (Ed), proceedings of the 12th International Workshop on Logic Programming Environments (WLPE 2002), July 2002, Copenhagen, Denmark.\\Proceedings of WLPE 2002: \texttt{http://xxx.lanl.gov/html/cs/0207052} (CoRR)}

\begin{abstract}
Previous work in the area of tracing \clpfd\ programs mainly focuses
on providing information about control of execution and domain
modification. In this paper, we present a trace structure that
provides information about additional important aspects. We
incorporate explanations in the trace structure, i.e.\ reasons for why
certain solver actions occur. Furthermore, we come up with a format
for describing the execution of the filtering algorithms of global
constraints. Some new ideas about the design of the trace are also
presented. For example, we have modeled our trace as a nested block
structure in order to achieve a hierarchical view. Also, new ways
about how to represent and identify different entities such as
constraints and domain variables are presented.
\end{abstract}

\section{Introduction}\label{sect:intro}
In this paper, we present new ideas in the area of tracing \clpfd\ 
programs. New trends such as tracing global constraints and explaining
solver actions are investigated. Since these issues are still in their
development stages and not systematically used, we have not tried to
create a universal trace model. Instead, we focus on \clpfd\ and in
particular the \clpfd\ solver of \sicstus\ Prolog. 




The idea of tracing constraint programs originates from
Meier~\cite{bib:mei95}. Also, the search-tree visualization tool in
CHIP~\cite{bib:sa00} uses a trace log to be able to display
propagation events. More current work is carried out by Langevine et
al.\ who presented a \clpfd\ trace model in~\cite{bib:lddj01a}. Their
work is naturally influenced by the work of
Ducass\'{e}~\cite{bib:duc99} on Opium.

Previous work such as the above mainly focuses on providing
information about control of execution and domain modification. Of
course, this information is needed, but if nothing else is provided,
important information about the execution is lost. We have looked at
the execution of a \clpfd\ program from the following perspectives:
\begin{enumerate}
\item\label{l11} Control of Execution; information about posting new
constraints, waking up delayed constraints, constraint
entailment/failure.
  
\item\label{l12} Domain Modification; information about variable
domain narrowing (pruning).
  
\item\label{l13} Declarative Aspect; information about why certain actions are
taken by the solver and in which context they occur. 
  
\item\label{l14} Procedural Aspect; information about filtering
algorithms and their entities.
\end{enumerate}

We build the trace of events as a nested block structure. This means
that most of the trace events appear inside a block surrounded by a
starting event and an ending event. One example of this is the
posting of a new constraint, which gives rise to a block surrounded by
the trace events \verb+begin_new_ctr+ and \verb+end_new_ctr+. Inside
this block, trace events describing the posting of demon(s) for the
new constraint and initial prunings are generated. Building the trace
like this has several advantages. First of all, it makes it easier to
get a structured overview of the trace. Second, similar actions may be
grouped together inside the same block. Furthermore, extra information
about the action(s) taking place may be added to the opening and
closing events as well as through added events inside the block.

We identify variables and other entities by their context, i.e.\ their
positions in the constraints, and not only by some identifier. This is
useful when one needs to distinguish a specific occurrence among
several of the same variable. In addition to that, it makes it
possible to also clearly identify entities such as integers, atoms and
compound terms.

Also, corresponding to point~\ref{l13} above, we provide
\dfn{explanations}~\cite{bib:jus01} for why certain actions
are taken by the solver. These explanations are incorporated in the
trace by specific explanation trace events. Actions for which
explanations are provided include domain narrowing, failure, demon
enqueuing and internal constraint posting. We extend previous work by
also providing compact explanations for global constraints. By doing
this, we obtain a stronger link between the consistency rules of the
global constraints and the trace.

Moreover, corresponding to point~\ref{l14} above, we provide trace
events containing detailed information about the filtering algorithms
of global constraints. This concerns identifying and presenting the
different entities inside the algorithms. This too increases the link
between the global constraints and the trace. However, this is more
related to the procedural aspect.

An example of a simple \clpfd\ program in \sicstus\ Prolog is shown in
Ex.~\ref{ex:traceme}. It will be used throughout the paper.  Line
\verb+(2)+ initializes the domains of the included variables.  Lines
\verb+(3)+ and \verb+(4)+ post the well-known constraints
\verb+all_different/1+\footnote{This is the naive version of
  \texttt{all\_different}, achieving the same pruning as a set of
  pairwise inequality constraints.} and \verb+element/3+. Line
\verb+(5)+ calls \sicstus's builtin search predicate with a
\verb+leftmost+\footnote{Leftmost variable, smallest value first.}
search strategy.

\begin{ex05}[h]
\begin{verbatim}
trace_me :-                                                           (1)
         domain([X,Y,V1,V2],1,6),                                     (2)
         all_different([X,Y,3,V1,8,V2])                               (3)
         element(X,[2,4,3,8],Y),                                      (4)
         labeling([leftmost],[X,Y,V1,V2]).                            (5)
\end{verbatim}
\caption{A simple \clpfd\ Program.}
\label{ex:traceme}
\end{ex05}

In the following, Sect.~\ref{sect:model} presents a \clpfd\ execution
model. This is needed in order to be able to present the trace
structure in its relevant context. Sect.~\ref{sect:trace} presents the
different trace events. First, some design choices concerning how to
represent different entities such as constraints and variables are
introduced. Following that, the trace events are presented in chunks;
trace events corresponding to the same perspective, as noted above,
are presented together. After that, a short description of the
implementation made in \sicstus\ Prolog is
given. Sect.~\ref{sect:relwork} discusses possible areas of
usage. Finally, Sect.~\ref{sect:concl} concludes the paper.
\section{Execution Model}\label{sect:model}
This section presents an execution model of a \clpfd\ kernel; the
different data structures it contains, and how these interact. By
doing this, we can later present the trace structure in its relevant
context. The model is influenced by the constraint solvers of
\sicstus\ Prolog~\cite{bib:coc97} and Choco~\cite{bib:lab00}.
\subsubsection{Data Structures.}
The data structures contain lists and queues for storing information
regarding constraints, variables, demons and propagation events.

Each \dfn{constraint} is associated with a list of domain variables,
\verb+VList+, and a list of demons, \verb+DList+. \verb+VList+
contains the variables that occur in the constraint. \verb+DList+
contains the demons that are responsible for removing inconsistent
values from the variables in \verb+VList+.

Each \dfn{variable} is associated with a domain, \verb+Domain+,
containing the values the variable may take. These domains are
narrowed by propagation events specified below. Furthermore, each
variable is associated with a list of constraints, \verb+CList+, which
it is involved in. As soon as the domain of a variable changes, one or
more of these constraints may wake up and activate their demons.

Each \dfn{demon} contains a \dfn{wake-up condition}. When the wake-up
condition is fulfilled, the demon's \dfn{filtering algorithm} is
activated. This filtering algorithm may exclude some values from the
domains of some variables. A filtering algorithm consists of a set of
\dfn{methods}, \mtt{\{M_{1},\ldots,M_{n}\}}, where each \mtt{M_{i}}
corresponds to some consistency rule of the constraint that the demon
belongs to. If some method, \mtt{M_{i}}, notices that the domains of
some variables are not consistent with the constraint, \mtt{M_{i}}
generates propagation events which specify domain narrowings on these
variables.

The \dfn{propagation events} have the form \verb+X in_set S+, which
constrains the variable \verb+X+ to be a member of the set \verb+S+ of
integers.

Two \dfn{global lists} are used for storing current constraints and
variables in the system; \verb+ConstraintList+ and
\verb+VariableList+. Furthermore, two \dfn{global queues} are needed
to manage the actions taken by the kernel. \verb+ReadyQueue+ contains
the demons that are about to wake up. \verb+PropagationQueue+ contains
the propagation events that have been created by active demons.

Fig.~\ref{fig:kernel} shows the different data structures and how they
are connected to each other. The state is taken from
Ex.~\ref{ex:traceme}, just before the execution of line
\verb+(4)+. \verb+C1+ and \verb+C2+ are the \verb+all_different/1+ and
\verb+element/3+ constraints respectively.  Detailed information about
variables \verb+X+ and \verb+Y+ as well as about constraint \verb+C2+
is shown in three boxes.

\figbegin{h} 
\begin{center}\mbox{\psfig{file=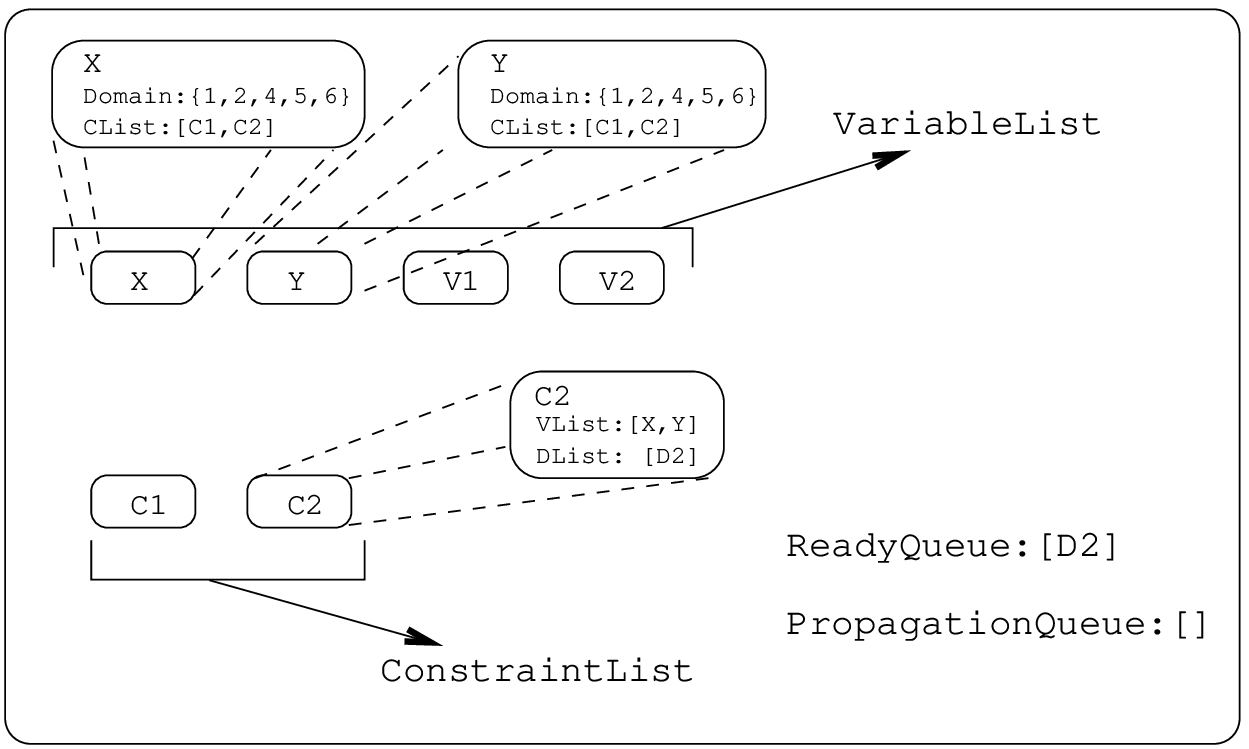,width=9cm}}\end{center}
\figend{fig:kernel}{Kernel state for Ex.~\ref{ex:traceme}. Just before 
the execution of line \texttt{(4)}.}    
\subsubsection{Services.}
Our \clpfd\ kernel provides the following services:
\begin{itemize}
  
\ttitem{connectCtr(C)}, adds the constraint \verb+C+ to the kernel.
This includes adding \verb+C+ to \verb+ConstraintList+, adding the
variables in \verb+C+ to \verb+VariableList+, \verb+adding+ \verb+C+
to the \verb+CList+ of all its variables, creating the
demons\footnote{Our trace structure supports the use of more than one
demon for each constraint. However, since the examples in this paper
are created using \sicstus\ Prolog, and it currently only associates
one demon with each constraint, more than one demon per constraint
will never be shown.} associated with \verb+C+ and running each
demon's filtering algorithm for initial propagation.
  
\ttitem{disconnectCtr(C)}, removes the constraint \verb+C+ from the
kernel. This is done when the constraint is entailed, i.e.\ when
\verb+C+ is true no matter what values its variables take.

\ttitem{enqueueCtr(C,E)}, puts last in \verb+ReadyQueue+ the demons
associated with the constraint \verb+C+ that are to wake up due to the
propagation event \verb+E+. If any of these demons are already in
\verb+ReadyQueue+, it will not be enqueued again.

\ttitem{dispatchCtr()}, dequeues the demon that is first in
\verb+ReadyQueue+ and activates its filtering algorithm.

\ttitem{enqueueEvent(E)}, puts \verb+E+ last in
\verb+PropagationQueue+.

\ttitem{dispatchEvent()}, dequeues the event \verb+E+ that is first in
\verb+PropagationQueue+, performs the narrowing on some domain
variable \verb+X+ it specifies, and calls \verb+enqueueCtr(C,E)+ for
all constraints in \verb+X+'s constraint list.
\end{itemize}
If the functions \verb+dispatchEvent()+ and \verb+dispatchCtr()+ can
both be executed, the former will have higher priority.
\subsubsection{The Active System: An Example.}
The easiest way to see how the above described data structures
interact is by looking at an example. Assume that the state of the
kernel is as in Fig.~\ref{fig:kernel}. There is no propagation event
to be performed, but the demon \verb+D2+ of constraint \verb+C2+, the
\verb+element/3+ constraint, should be woken up. The kernel calls the
function \verb+dispatchCtr()+, which removes \mtt{D2} from
\verb+ReadyQueue+ and activates its filtering algorithm. A first
method, \verb+M1+, generates a propagation event, \verb+E1+, which
restricts \verb+X+ to take a value inside the range \verb+1..4+.
Equally, a second method, \verb+M2+, generates a propagation event,
\verb+E2+, which restricts \verb+Y+ to take a value inside the set
\verb+{2,3,4}+. Also, a third method, \verb+M3+, generates a
propagation event, \verb+E3+, which restricts \verb+X+ from taking the
value \verb+4+\footnote{Since the value \texttt{8} on position
\texttt{4} in the list is not in the domain of \texttt{Y}.}. When the
filtering algorithm of \verb+D2+ has finished, the kernel notices that
\verb+PropagationQueue+ has changed to contain the propagation events
\verb+E1+, \verb+E2+ and \verb+E3+.  The kernel calls the function
\verb+dispatchEvent()+ three times which does the following each time:
\begin{itemize}
\item The first propagation event in \verb+PropagationQueue+ is
dequeued.
\item The narrowing of the variable specified by the propagation event
is performed by the kernel.
\item The demons of the constraints \verb+C1+ and \verb+C2+, that are
supposed to wake up due to the narrowing performed, are put in
\verb+ReadyQueue+ by calling the function \verb+enqueueCtr()+
twice. 
\end{itemize}
When \verb+PropagationQueue+ is empty, the kernel starts over by
calling the function \verb+dispatchCtr()+ again. This is repeated
until both queues are empty. When this is the case, the control of
execution is given back to the Prolog level. In our case, this means
that labeling takes place which will fix a variable and therefore
trigger more propagation.

\section{Trace Structure}\label{sect:trace}
In this section, the trace structure and the information it contains
is presented. We start by introducing our ideas about how to identify
the different entities in a \clpfd\ program such as constraints and
variables. Following that, the trace events are presented together
with some examples. Finally, our implementation in \sicstus\ Prolog is
described.
\subsection{Representing and Identifying Entities in \clpfd\
Programs}\label{sect:tracerep}
\subsubsection{Identifiers.} 
The \dfn{id of a constraint} is a Prolog atom created by prepending
the atom \verb+ctr_+ to the source code functor and appending a
unique\footnote{Unique for the specific type of constraint.} number to
this. The \dfn{id of a variable} is a Prolog atom of the form
\verb+fdvar_N+ where \verb+N+ is a unique number. For example, the
following identifiers are produced by Ex.~\ref{ex:traceme} where
\verb+C0+, \verb+C1+ and \verb+C2+ are the constraints posted on lines
\verb+(2)+, \verb+(3)+ and \verb+(4)+ respectively: {\tracesize
\begin{quotetab}
  \{\=\mtt{X \mapsto fdvar\_1,Y \mapsto fdvar\_2,V1 \mapsto
    fdvar\_3,V2 \mapsto fdvar\_4,} \\
  \>\mtt{C0 \mapsto ctr\_domain\_1,C1 \mapsto
    ctr\_all\_different\_1,C2 \mapsto ctr\_element\_1}\}
\end{quotetab}
}
\subsubsection{Constraint Representation.}
\label{sect:conrep}
A constraint is represented similarly to its source code
representation. It has the same functor atom and structure, with all
variables replaced by their given identifiers. In Ex.~\ref{ex:traceme}
for instance, the constraint {\tracesize
\begin{quote}
\verb+all_different([X,Y,3,V1,8,V2])+ 
\end{quote}}
is represented as {\tracesize
\begin{quote}
\verb+all_different([fdvar_1,fdvar_2,3,fdvar_3,8,fdvar_4])+.
\end{quote}}
\subsubsection{Location of Variables.}
It is not enough to be able to identify variables by their
identifiers. While this allows to distinguish different variables, it
will not make it possible to, within one constraint, uniquely
determine a specific occurrence among several of the same variable.

We have solved this with what we call a \dfn{path}. A path uniquely
determines a position or a list of positions in a constraint. When
referencing arguments of a constraint, for example when displaying a
variable's domain, a path referring to that variable is given instead
of its id. This path must always have a context, a corresponding
constraint, to be interpreted in.

A path is represented as a Prolog list containing (lists of) fixed
values or other paths. The first element of a path identifies the
topmost position in the corresponding constraint. For example, any
path referring to some position(s) in the \verb+element/3+ constraint
in Ex.~\ref{ex:traceme} has one of the integers 1, 2, 3 or one of the
lists \verb+[1,2]+, \verb+[1,3]+, \verb+[2,3]+, \verb+[1,2,3]+ as
first element. Lines a and b in Table~\ref{ex:path1} show two of
these in paths with only one element and the position(s) they refer
to.

The i+1:st element of a path refers to some position(s)
\standout{inside} whatever the i:th element refers to. This means that
the position(s) the i:th element refers to must be a compound term or
a list of compound terms. Line c in Table~\ref{ex:path1} illustrates
this. This example also introduces one more symbol in the path term;
the \verb+#+ sign. It is used whenever the term(s) we refer to is
inside a list (a non-list compound term would not have a preceding
\verb+#+ sign).

Some more syntax is needed in order for the path to be more expressive
and compact. First of all, we will use \verb+[*]+ to express ``all
positions'' of a compound term. It may also be useful to identify
\standout{almost} all positions in a compound term. We will use the
\verb+\+ sign for this purpose, denoting set subtraction\footnote{In
our case, the sets are lists.}. Line d in Table~\ref{ex:path1}
illustrates this.

In addition to the above, we may apply functions to the entities
referred to by a path. \verb+P/F+ has the meaning that the function
\verb+F+ is applied to each entity referred to by the path
\verb+P+. Possible values for \verb+F+ are, among others, \verb+min+,
\verb+max+ and \verb+length+. Table~\ref{ex:path2} shows some examples
of this. If \verb+P+ refers to a list of positions of size \mexp{n},
\verb+P/F+ returns a list of values of size \mexp{n}. However, one
element lists are simplified as shown on lines a and b in
Table~\ref{ex:path2}.

\tabbegin{h}
\caption{The positions marked with gray are the positions referred to
by the corresponding path.}
\begin{tabular}{lll|lll}
\hlineskip
& \tabhead{Path} & \tabhead{Position(s)} & & \tabhead{Path} &
\tabhead{Position(s)} \\ 
\hlineskip
a. & \verb+[[1,3]]+ & \texttt{element(\greybox{X},[2,4,3,8],\greybox{Y})} &
c. & \verb+[2,#[1,2]]+ &
\texttt{element(X,[\greybox{2},\greybox{4},3,8],Y)}  \\
b. & \verb+[2]+ & \texttt{element(X,\greybox{[2,4,3,8]},Y)} &
d. & \verb+[2,#[*]]\[2,#1]+ &
\texttt{element(X,[2,\greybox{4},\greybox{3},\greybox{8}],Y)} \\
\hlineskip
\end{tabular}
\label{ex:path1}
\tabend
\tabbegin{h}
\caption{Functions applied to paths. The state of the variables is as
  in Fig.~\ref{fig:kernel}, i.e.\ before the execution of line
  \texttt{(4)} in Ex.~\ref{ex:traceme}.}
\begin{tabular}{llll}
\hlineskip
& \tabhead{P/F} & \tabhead{Constraint} & \tabhead{Value(s)} \\ 
\hlineskip
a. & \verb+[1]/length+ & \verb+all_different([X,Y,3,V1,8,V2])+ &
\verb+6+ \\
b. & \verb+[1]/min+ & \verb+element(X,[2,4,3,8],Y)+ & \verb+1+ \\
c. & \verb+[[1,3]]/max+ & \verb+element(X,[2,4,3,8],Y)+ & \verb+[6,6]+ \\
\hlineskip
\end{tabular}
\label{ex:path2}
\tabend
\subsection{Trace Events}
Now, it is time to present the trace events. They are organized in
chunks, each chunk containing trace events for a specific purpose. The
chunks of trace events are presented in tables followed by examples
generated by Ex.~\ref{ex:traceme}.

All trace events have an \verb+EventId+ attribute, an increasing
number. We will often identify them by the value of this attribute and
simply call it \dfn{id}. The \verb+EventId+ is always the first
attribute of the Prolog fact. Furthermore, when we talk about a whole
block of trace events, we mean the opening and closing trace events of
the block and all trace events in between. For example, a \verb+prune+
block includes the trace events \verb+begin_prune+ and
\verb+end_prune+, and all trace events between these.

Within our illustrations of Ex.~\ref{ex:traceme}, since the constraint
representations occur very frequently, we will use the following short
notations: {\tracesize
\begin{itemize}
\item \verb+ALLDIFF = all_different([fdvar_1,fdvar_2,3,fdvar_3,8,fdvar_4])+
\item \verb+ELEMENT = element(fdvar_1,[2,4,3,8],fdvar_2)+
\end{itemize}
} Also, when mentioning variables, we will use the variable
name from the source code of Ex.~\ref{ex:traceme}, even if the trace
events contain the given identifier.
\subsubsection{Control of Execution.}
In this section, trace events describing the control of execution are
presented. This concerns things like new constraint posting, demon
enqueuing and demon wakening. The trace events that describe these
actions are presented in Table~\ref{ev:exec}.

\tabbegin{h}
\caption{Control of execution trace events.}
\evtabbegin{2.7}{2.1}{4.8}

\verb+begin_new_ctr+ & \verb+EventId+, \verb+ConstraintId+,
\verb+Constraint+ & Generated when a new constraint is to be added to
the system, i.e.\ when entering the \verb+connectCtr()+
function. \verb+ConstraintId+ and \verb+Constraint+ are respectively
the constraint's id and internal representation as explained above. \\
\hlineskip

\verb+end_new_ctr+ & \verb+EventId+, \verb+Result+ & Generated when
demons for the constraint have been created and possible initial
propagation has finished, i.e.\ just before exiting the
\verb+connectCtr()+ function. \verb+Result+ denotes the result of
adding the new constraint. \mtt{Result \in
\{fail,delay,entail\}} \\\hlineskip

\verb+new_demon+ & \verb+EventId+, \verb+DemonId+\footnote{Identifiers
  for demons are generated similar to identifiers for constraints.},
\verb+DemonType+, \verb+WakeConds+, \verb+Constraint+ & Generated when
a new demon is added to the system. This happens inside the
\verb+connectCtr()+ function. \verb+DemonType+ is a descriptive atom
of the demon. \verb+WakeConds+ is a list of pairs
(\verb+VPath+,\verb+Condition+) where \verb+VPath+ is a path to some
variable(s) and \mtt{Condition \in \{min,max,minmax,val,dom\}}.  \\
\hlineskip

\verb+push_demon+ & \verb+EventId+, \verb+DemonId+, \verb+DemonType+ &
Generated when a wake-up condition for at least one variable has been
fulfilled. This corresponds to the function \verb+enqueueCtr()+.\\
\hlineskip

\verb+begin_wake_demon+ & \verb+EventId+, \verb+DemonId+,
\verb+DemonType+ & Generated when a demon is to wake up for
propagation. This corresponds to entering the \verb+dispatchCtr()+
function, after dequeuing the topmost demon from \verb+ReadyQueue+.\\
\hlineskip

\verb+end_wake_demon+ & \verb+EventId+, \verb+Result+ & Generated when
a demon has finished all propagation. This corresponds to exiting the
\verb+dispatchCtr()+ function, after the filtering algorithm has
finished.\\
\hlineskip \evtabend
\label{ev:exec}
\tabend

Lines \verb+(3)+ and \verb+(4)+ in Ex.~\ref{ex:traceme} post two
constraints to the system. This gives rise to the following trace
events to be created\footnote{There are several gaps, identified by
`\ldots', since we only present the trace events relevant to the
control of execution at this stage.}:
{\tracesize
\begin{verbatim}
begin_new_ctr(9,ctr_all_different_1,ALLDIFF)
  new_demon(10,ctr_all_different_1,all_different_1,[[1,#[*]]-val],
            ALLDIFF)
  begin_wake_demon(11,ctr_all_different_1,all_different_1)
  ...
  end_wake_demon(21,delay)
end_new_ctr(22,delay)
begin_new_ctr(23,ctr_element_1,ELEMENT)
  ...
  begin_new_ctr(25,ctr_in_1,fdvar_1 in 1..4)
  ...
  end_new_ctr(32,entail)
  new_demon(33,ctr_element_1,element_3,
            [[1]-dom,[2,#[*]]-minmax,[3]-minmax],ELEMENT)
  begin_wake_demon(34,ctr_element_1,element_3)
  ...
  end_wake_demon(57,delay)
end_new_ctr(58,delay)
\end{verbatim}
  } The first five trace events are created due to the posting of the
constraint \verb+all_different/1+. This gives rise to the creation of
a new demon (id \verb+10+) and the wakening of that demon (id:s
\verb+11+ and \verb+21+). The last one of these trace events (id
\verb+22+) contains the result \verb+delay+, meaning that the
constraint will not wake up again until the state of at least one of
its variables changes. Moving on, the \verb+element/3+ constraint is
introduced in much the same way. The difference is the creation of an
internal \verb+in/1+ constraint (id:s \verb+25+ and \verb+32+) to
ensure that \verb+X+ takes a value in the range \verb+1..4+.
\subsubsection{Domain Modification.}
Now, we introduce trace events describing domain modification, i.e.\
pruning of variables. Such trace events are presented in
Table~\ref{ev:dmod}.  
\tabbegin{h}
\caption{Domain modification trace events.} 
\evtabbegin{2.1}{1.8}{3.9}

\verb+begin_prune+ & \verb+EventId+, \verb+Intention+,
\verb+Constraint+ & Generated when some pruning is about to occur,
corresponds to actions inside the \verb+dispatchCtr()+
function. \verb+Intention+ is a Prolog fact with information about the
intended pruning\footnote{Since some values might have been removed
already, the actual values removed may be different from the intended
pruning.}.\\\hlineskip

\verb+end_prune+ & \verb+EventId+, \verb+Result+ & Generated when some
pruning has been carried out. \verb+Result+ denotes the result of all
domain narrowings performed inside the block. \mtt{Result = succeed}
iff all prunings inside the block was successful, otherwise
\mtt{Result = fail}.\\\hlineskip

\verb+prune+ & \verb+EventId+, \verb+PrunedVars+, \verb+Pruning+,
\verb+Constraint+, \verb+Result+ & Generated when some domain
narrowing occurs. This corresponds to the \verb+enqueueEvent()+
function. \verb+PrunedVars+ is a path referring to the positions of
the pruned variables. \verb+Pruning+ is a Prolog fact with information
about the actual pruning. \verb+Result+ denotes the result,
\verb+fail+ or \verb+succeed+, of the pruning. \\\hlineskip

\verb+before_prune+, \verb+after_prune+ & \verb+EventId+,
\verb+PrunedVars+, \verb+Domains+, \verb+Constraint+ & These are
generated before (after) a \verb+prune+ trace event. They contain
domain information about the pruned variables before (after) value
removal. \verb+Domains+ is a list of domains in the \sicstus\ Prolog
format\footnote{For example, \texttt{[[3|5],[7|7]]} denotes the set
\texttt{\symbol{`\{}3,4,5,7\symbol{`\}}}, see the \clpfd\ section
of~\cite{bib:sicstus} for details.}. \\\hlineskip

\verb+fail+ & \verb+EventId+, \verb+Constraint+ & Generated when
inconsistency is noticed due to some non-fulfilled necessary condition
of a constraint. \\\hlineskip

\evtabend
\label{ev:dmod}
\tabend

For Ex.~\ref{ex:traceme}, the following trace events are added due to
the posting of the \verb+all_different/1+ constraint on line
\verb+(3)+:
{\tracesize
\begin{verbatim}
begin_prune(14,remove_value(3),ALLDIFF)
  ...
  before_prune(16,[1,#[1,2,4,6]],[[[1|6]],[[1|6]],[[1|6]],[[1|6]]],
               ALLDIFF)
  prune(17,ctr_all_different_1,[1,#[1,2,4,6]],remove_value(3),ALLDIFF,
        succeed)
  after_prune(18,[1,#[1,2,4,6]],[[[1|2],[4|6]],[[1|2],[4|6]],
                                 [[1|2],[4|6]],[[1|2],[4|6]]],ALLDIFF)
end_prune(19,succeed)
\end{verbatim}
} 
These will occur between the trace events with id:s \verb+11+ and
\verb+21+, the wakening of the \verb+all_different/1+ demon. The first
trace event contains information about the intended domain
narrowing. In this case the value \verb+3+ is to be removed from some
variables. The path to these variables is shown in the following three
trace events. The first and last of these (id:s \verb+16+ and
\verb+18+) contain domain information about the variables before and
after pruning. The middle trace event (id \verb+17+) contains
information about the actual pruning. Finally, the trace event with id
\verb+19+ contains the result of the pruning. Since it did not produce
any empty domains, the result is \verb+succeed+.

The posting of the \verb+element/3+ constraint generates the following
trace events containing information about pruning the variable
\verb+X+:
{\tracesize
\begin{verbatim}
begin_prune(50,remove_values([[4|4]]),ELEMENT)
  ...
  before_prune(52,[1],[[[1|2],[4|4]]],ELEMENT)
  prune(53,ctr_element_1,[1],remove_value(4),ELEMENT,succeed)
  after_prune(54,[1],[[[1|2]]],ELEMENT)
end_prune(55,succeed)
\end{verbatim}
}
\subsubsection{Explanations.}\label{sect:expl}
The trace events presented in this section provide
\dfn{explanations}~\cite{bib:jus01} for other trace events, i.e.\
reasons for why certain actions are taken by the solver. These trace
events are presented in Table~\ref{ev:expl}.

\tabbegin{h} 
\caption{Explanation trace events. All but
\texttt{push\_demon\_because} contain the same type of information.}
\evtabbegin{3}{2.1}{5.1}

\verb+push_demon_because+ & \verb+EventId+, \verb+Variables+,
\verb+Condition+, \verb+Constraint+ & Generated before a
\verb+push_demon+ trace event. \verb+Condition+ is a property
fulfilled by the variables referred to by the path
\verb+Variables+. \mtt{Condition \in \{min,max,minmax,val,dom\}}.
\\\hlineskip

\verb+prune_because+, \verb+fail_because+, \verb+new_ctr_because+ &
\verb+EventId+, \verb+ConstraintId+, \verb+Explanation+,
\verb+Constraint+ & Generated before the
corresponding\footnote{Trivially, a \texttt{prune\_because} trace
event is for example generated before some trace events describing
domain narrowing.} trace events.  \verb+Explanation+ is an explanation
or a reason for why the corresponding action occurs.\\\hlineskip

\evtabend
\label{ev:expl}
\tabend

First of all, we will make clear what we mean with explanations: An
explanation \mexp{e} for some kernel action \mexp{a} is a logical
formula \mexp{e} such that \mexp{e \Rightarrow a}, i.e.\ if \mexp{e}
holds, \mexp{a} will occur. In Prolog, we represent an explanation
with a pair, \verb+N-CondList+, where \verb+N+ is an integer and
\verb+CondList+ is a list of \verb+cond(M,Pa,PL)+ terms.  An
explanation \mexp{e =} \verb+N-CondList+ holds iff at least \verb+N+
of the \verb+cond/3+ terms in \verb+CondList+ hold. Each \verb+cond/3+
term contains an integer \verb+M+, a path \verb+Pa+ and a list of
properties \verb+PL+. The property on index \mexp{i} in \verb+PL+ is
associated with the entity on index \mexp{i} in the list of positions
referred to by \verb+Pa+. For example, assume that \verb+Pa+ $=$
\verb+[[1,3]]+, \verb+PL+ $=$ \verb+[eq(2),neq(3)]+ and that \verb+Pa+
is associated with the \verb+element/3+ constraint in
Ex.~\ref{ex:traceme}. Then \verb+X+ is associated with \verb+eq(2)+
and \verb+Y+ is associated with \verb+neq(3)+. When \verb+PL+ contains
a single property, it is associated with each entity referred to by
\verb+Pa+. Also, the list structure is skipped and \verb+PL+ will
denote the value of the single property. Possible properties in
\verb+PL+ are, among others, \verb+eq(I)+, \verb+neq(I)+,
\verb+inset(S)+ and \verb+notinset(S)+ where \verb+I+ is an integer
and \verb+S+ is a finite set of integers. A \verb+cond/3+ term holds
iff at least \verb+M+ of the entities referred to by \verb+Pa+ fulfill
their respective properties.


Now, let us give an illustrative example. Assume that 
\begin{quote}
\mexp{e =} \verb+1-[cond(1,[1,#3],eq(3))]+
\end{quote}
explains the solver action \mexp{a}. Also, assume that \mexp{e} is
associated with a pruning generated due to the \verb+all_different/1+
constraint in Ex.~\ref{ex:traceme}. This gives us the information that
\mexp{a} occurs since at least one of the \verb+cond/3+ terms
holds. For the only one, it holds since at least one of the entities
referred to by \verb+[1,#3]+ can take the value \verb+3+,
\verb+eq(3)+. In our case, this is the position marked with gray in
\texttt{all\_different([X,Y,\greybox{3},V1,8,V2])}. In order for
\mexp{a} not to occur, the least thing one must do is to change the
value on that position to some value distinct from \verb+3+. Actually,
this explanation occurs in the trace event {\tracesize
\begin{verbatim}
prune_because(15,ctr_all_different_1,1-[cond(1,[1,#3],eq(3))],ALLDIFF).
\end{verbatim}
} 
Hence, \mexp{e} is an explanation for removing the value \verb+3+ from
all variables in the \verb+all_different/1+ constraint.

The following are some more explanation trace events from
Ex.~\ref{ex:traceme}:
{\tracesize
\begin{verbatim}
new_ctr_because(24,ctr_element_1,1-[cond(1,[2]/length,eq(4))],ELEMENT)
...
prune_because(51,ctr_element_1,1-[cond(1,[3],notinset([[8|8]]))],ELEMENT)
\end{verbatim}
} The first one is generated before the \verb+begin_new_ctr+
trace event with id \verb+25+. It contains an explanation for posting
the \verb+X in 1..4+ constraint inside the creation of the
\verb+element/3+ constraint. The explanation is that the length of the
list on the second position in the constraint is \verb+4+.

Similarly, the id of the last one tells us that it is generated before
the \verb+before_prune+ trace event with id \verb+52+. It contains an
explanation for pruning the variable \verb+X+. This too occurs inside
the creation of the \verb+element/3+ constraint. The explanation is
that the third argument of the constraint, the variable \verb+Y+, does
not intersect with the set consisting of the single value
\verb+8+. Due to this, \verb+X+ cannot take the value \verb+4+.

For some solver actions, there are several different reasons for why
they occur. In order to show this, we need to introduce one more
example.

\begin{ex05}
\begin{verbatim}
explain_me :-                                                         (1)
         domain([X1],1,6),                                            (2)
         domain([X2,X3,X4,X5],1,2),                                   (3)
         all_distinct([X1,X2,X3,X4,X5]).                              (4)
\end{verbatim}
\caption{A failing \texttt{all\_distinct/1} constraint.}
\label{ex:explainme}
\end{ex05}

The \verb+all_distinct/1+ constraint on line \verb+(4)+ in
Ex.~\ref{ex:explainme} uses the complete filtering algorithm of
R\'egin~\cite{Reg94}. When, during the posting of that constraint, its
demon is woken up, it will produce a failure since the variables
\verb+X2+, \verb+X3+, \verb+X4+ and \verb+X5+ are all in the range
\verb+1..2+. For this failure, we generate the following explanation:
\begin{quote}
\verb+1-[cond(all,[1,#[2,3,4,5]],inset([[1|2]]))]+
\end{quote}
where \verb+all+ is a keyword for identifying all entities referred to
by the associated path. By looking at this explanation, we conclude
that, in order for the failure not to occur, we need to enlarge the
domains of any two of the variables \verb+X2+, \verb+X3+, \verb+X4+
and \verb+X5+ with at least two distinct values. We see this since
there are $4$ variables taking values in a set of integers of size $2$
and $4-2=2$.
\subsubsection{Filtering Algorithms.}
The trace events presented in this section describe the execution of
the filtering algorithms associated with the demons. Since each
filtering algorithm is built up by a set of methods, information about
each active method is needed. Table~\ref{ev:filter} presents the
relevant trace events.  \tabbegin{h}
\caption{Filtering algorithms trace events.}      
\evtabbegin{2}{1.8}{3.8}

\verb+begin_method+ & \verb+EventId+, \verb+MethodName+ & Generated
when a demon's filtering algorithm activates a specific
method. \verb+MethodName+ is a Prolog atom with a name corresponding
to the consistency rule of the associated constraint.\\\hlineskip

\verb+end_method+ & \verb+EventId+, \verb+Result+ & Generated when a
specific method has finished. \verb+Result+ denotes the result of the
prunings carried out inside the block. Similar to the \verb+Result+
attribute in the \verb+end_prune+ trace event.\\\hlineskip

\verb+info_method+ & \verb+EventId+, \verb+InfoName+, \verb+Info+,
\verb+Constraint+ & Generated inside a specific method. Contains more
detailed information about entities involved in the filtering
algorithm. \verb+InfoName+ is a Prolog atom with a descriptive name of
the provided information. \verb+Info+ is a pair \verb+P-E+ containing
the extra information, \verb+P+ is a path referring to the entity
\verb+E+. \\\hlineskip

\evtabend
\label{ev:filter}
\tabend

Each \verb+prune+ block is surrounded by a \verb+method+ block, and
possibly preceded by an \verb+info_method+ trace event. For the first
example concerning domain modification, the following trace events are
added:
{\tracesize
\begin{verbatim}
begin_method(12,propagate_ground_variable)
  info_method(13,ground_variable,[1,#3]-3,ALLDIFF)
  ...
end_method(20,succeed)
\end{verbatim}
} By looking at the id:s, we see that they are generated
before and after the \verb+prune+ block describing the removal of the
value \verb+3+ from the variables \verb+X+, \verb+Y+, \verb+V1+ and
\verb+V2+. The \verb+begin_method+ trace event contains a name of the
active method. In this case, \verb+propagate_ground_variable+ denotes
the method that removes a ground value from all non-ground domain
variables in the \verb+all_different/1+ constraint. Following that,
the \verb+info_method+ trace event contains more detailed information
about the actual entities involved in the method. More specifically,
it tells us that the ground variable is on the position referred to by
\verb+[1,#3]+ and its value is \verb+3+. Finally, the
\verb+end_method+ trace event contains the result, \verb+succeed+, of
the whole \verb+method+ block. The result will be \verb+fail+ iff at
least one failure inside the block is detected, otherwise
\verb+succeed+.

The second example concerning domain modification, the one associated
with the \verb+element/3+ constraint, is surrounded by the following
\verb+method+ block:
{\tracesize
\begin{verbatim}
begin_method(49,prune_x)
...
end_method(56,succeed)
\end{verbatim}
}
\subsection{Implementation}\label{sect:impl}
The described trace structure has been implemented in a research
branch of the \clpfd\ solver in \sicstus\ Prolog version 3.9.

Each trace event is represented by an abstract datatype (ADT) in C.
Complex information, such as compound terms and lists, is represented
using C strings. Each ADT has creation and deletion primitives as well
as a primitive for converting the information to a Prolog fact. 

On the Prolog side, it is possible to create and access the trace
events through some predicates. For the creation, each trace event
(with two exceptions\footnote{The trace events \texttt{push\_demon}
and \texttt{push\_demon\_because} are created at a central place
within the C part of the kernel.}) have a corresponding constructor
predicate. For retrieving, a predicate
\verb|get_events(+Selector, -TraceEvents)| is
available. \verb+Selector+ is an integer, an atom or a compound term
specifying a set of trace events. \verb+TraceEvents+ is a list of
trace events corresponding to \verb+Selector+.


\section{Discussion}\label{sect:relwork}
Tracing \clpfd\ programs is still an open area. In our case, more
testing is needed in order to evaluate the trace model. This includes
building debugging tools, using the different kinds of information the
trace provides.

One interesting tool to build would be a search-tree visualizer such
as the Oz Explorer~\cite{bib:schulte97} by Schulte. Such a tool would
explore the information in the trace concerning the control of
execution aspect. Also, as was done for the Oz Explorer, one could
connect to it other tools using a plug-in scheme, displaying various
information concerning other aspects at the nodes of the
search-tree. An example of such a tool for the Oz Explorer is the
Constraint Investigator~\cite{bib:mueller00} by M\"uller.

Another tool, exploring the information concerning the domain
modification aspect, would be a variable domain visualizer. Domains
could for instance be visualized in a way similar to as
in~\cite{bib:car00}.

The information concerning the declarative aspect could be used by a
tool for solving over-constrained problems. Explanations would help to
point out the set of inconsistent constraints. Jussien and Ouis
presented a tool like this in~\cite{bib:jo01}.

For educational purposes, a tool that provides information about
filtering algorithms would be useful. Such a tool could connect the
consistency rules of the constraints with the different methods in the
corresponding filtering algorithms.

As a first use of the trace structure, we have implemented an
extension to the Prolog debugger in \sicstus. At the usual breakpoints
of the debugger, it is possible to ask queries about the trace and the
entities in it. The user may for example display certain trace events,
ask queries about domain variables and generate explanations for
specific domain narrowings.
\section{Conclusion}\label{sect:concl}
In this paper, we presented new ideas in the area of tracing \clpfd\ 
programs. In addition to providing information about control of
execution and domain modification, we also provide information
concerning the declarative and procedural aspects. We did this by
incorporating the idea of explanations in the trace, as well as coming
up with new ideas about tracing filtering algorithms. By doing this,
we have obtained a link between the consistency rules and the
filtering algorithms in constraint solving.

Furthermore, we presented some new ideas about how to present the
trace, both regarding the structure and the information contained.
This includes building the trace as a nested block structure, as well
as using a path for identifying the different entities in a \clpfd\ 
program.

Implementing the trace structure in a high-end system like \sicstus\
is not an easy task. This is especially true for the task of adapting
filtering algorithms to generate explanations. Sometimes, the added
overhead it meant to provide the information needed for sharp
explanations was too large. In this case, some \dfn{approximation
explanation}\footnote{An explanation that is true but not necessarily
the one that expresses the least dependency.} has been used. In the
long run, providing explanations probably means that filtering
algorithms need to be changed in order to, more naturally, provide the
necessary information.

\end{document}